\documentclass[twocolumn,showpacs,preprintnumbers,amsmath,amssymb]{revtex4}

\usepackage{graphicx}
\usepackage{dcolumn}
\usepackage{epsfig}
\usepackage{bm}
\usepackage{bbm}
\def\beq{\begin{equation}}
\def\eeq{\end{equation}}
\def\be{\begin{equation}}
\def\ee{\end{equation}}
\def\bea{\begin{eqnarray}}
\def\eea{\end{eqnarray}}

\begin{document}

\title{Duality in Left-Right Symmetric Seesaw Mechanism}

\author{E. Kh. Akhmedov$^{a,b}$ and M. Frigerio$^{c}$ }
\affiliation{
\vspace*{1mm}
$^a$ Physik Department T30, Technische Universit\"at M\"unchen,
James-Franck Stra\ss e, D-85748 Garching, Germany\\
$^b$ Kurchatov Institute,  Moscow, Russia\\
$^c$ Department of Physics, University of California, Riverside, 
CA 92521, USA }

\begin{abstract}
We consider type I+II seesaw mechanism, where the exchanges of both 
right-handed neutrinos and isotriplet Higgs bosons contribute to the 
neutrino mass. Working in the left-right symmetric framework and assuming 
the mass matrix of light neutrinos $m_\nu$ and the Dirac-type Yukawa couplings 
to be known, we find the triplet Yukawa coupling matrix $f$, which carries the 
information about the masses and mixing of the right-handed neutrinos. We show 
that in this case there exists a duality: for 
any solution $f$, there is a dual solution $\hat{f}=m_\nu/v_L-f$, where $v_L$ 
is the VEV of the triplet Higgs. Thus, unlike in pure type I (II) seesaw, 
there is no unique allowed structure for the matrix $f$. For $n$ lepton 
generations the number of solutions is $2^n$. We develop an exact analytic 
method of solving the seesaw non-linear matrix equation for $f$. 
\end{abstract}

 \hspace*{3cm} 

\maketitle

%%%%%%%%%%%%%%%%%%%%%%%%%%%%%%%%%%%%%%%%%%%%

{\it 1. Introduction. }
The discovery of neutrino oscillations in atmospheric, solar, reactor and 
long-baseline neutrino experiments \cite{expsummary} gave an unambiguous 
evidence for neutrino mass and mixing and spurred the activity in 
neutrino mass model building. Among the possible mechanisms of neutrino 
mass generation, arguably the simplest and most attractive one is the 
seesaw mechanism \cite{seesawI,seesawII}, which explains the smallness of 
the neutrino masses through the exchange of superheavy particles.
In addition, it has an elegant built-in mechanism of generation of the 
baryon asymmetry of the universe -- baryogenesis via leptogenesis 
\cite{leptog}. 

In its simplest version, the seesaw mechanism extends the standard electroweak 
model by assuming the existence of $SU(2)_L$-singlet right-handed (RH) 
neutrinos $N_{R}$ with a bare Majorana mass matrix $M_{R}$. The RH neutrino 
masses, not being protected by the electroweak symmetry, are naturally large 
-- at Planck or GUT scale. The masses of light neutrinos, produced in this 
(so-called type~I) seesaw mechanism, are \cite{seesawI} 
\be
m_\nu \simeq m_\nu^I = -m_D {M_R}^{-1} m_D^T\,,
\label{ss1}
\ee
where $m_D=vy$, $v\simeq 174$ GeV being the electroweak VEV and $y$ the Dirac-type Yukawa matrix describing 
the coupling of lepton doublets to $N_R$ and the standard Higgs doublet. 
Alternatively, the masses of light neutrino can be generated through their 
interaction with a heavy $SU(2)_L$-triplet Higgs scalar $\Delta_L$, which 
can acquire an induced VEV $v_L$ through its coupling to the 
Higgs doublet. This (so-called type II) seesaw mechanism \cite{seesawII} 
yields 
\be
m_\nu=m_\nu^{II} = v_L f_L \sim (v^2/M_{\Delta})  f_L \,,
\label{ssII} 
\ee
where $f_L$ is the Yukawa coupling between the lepton doublets and 
$\Delta_L$, and $M_\Delta$ is the mass of the isotriplet Higgs. 

The RH neutrinos $N_R$, being electroweak singlets, are actually aliens to 
the standard electroweak model, though can be freely added to it. They are 
much more natural in models with left-right (LR) symmetry \cite{LR} 
or in $SO(10)$ GUT \cite{SO10}, which can be broken down to the standard 
model through the LR-symmetric stage. 
The LR symmetry provides a very natural explanation of the observed
maximal parity and C violation in low-energy weak interactions and is
therefore likely to be present at some level in the final theory.

In this class of models, the RH neutrinos 
are in doublets $l_R$ of $SU(2)_R$, and their mass is generated by the 
VEV $v_R$ of the $SU(2)_R$-triplet Higgs field $\Delta_R$ through the 
Majorana-like Yukawa coupling. LR symmetry then implies that there should 
also exist an $SU(2)_L$-triplet Higgs field $\Delta_L$, and the Yukawa 
couplings of the triplet Higgs scalars to leptons have the form 
\be
f_L 
\left(l_L^T C \,i\tau_2 \,\Delta_L \, l_L \right) 
+ f_R \left(l_L^{cT} C \,i\tau_2 \, \Delta_R \,l_L^c\right)\! 
+ h.c.\,,
\label{tripl}
\ee
where $l_L^c \equiv C \overline{l_R}^T$, $\Delta_{L,R}=\mbox{\boldmath 
$\tau$}{\bf \Delta}_{L,R}$ and   $C$ is the charge conjugation matrix. 
The Dirac-type Yukawa matrix $y$ comes from the coupling 
of the lepton doublets $l_L$ and $l_R$ with the bi-doublet Higgs field 
$\Phi$. In this class of models, 
light neutrino masses naturally get both type I and type II 
contributions, with $M_R\equiv v_R f_R$.

In addition to the gauge LR symmetry, the models of this kind are usually 
assumed to possess a discrete LR symmetry, which is broken at a scale 
$v_{LR}$ that may or may not coincide with the $SU(2)_R$ breaking scale 
$v_R$. There are essentially two ways in which the discrete LR symmetry 
can be introduced. 
In the first one, the left-handed and RH fermions are interchanged, 
while the Higgs fields undergo $\Delta_L \leftrightarrow \Delta_R^*$, $\Phi 
\leftrightarrow \Phi^\dagger$. This yields 
\be
f_L=f_R^*\,,\qquad\quad y=y^\dagger\,,
\label{LR1a}
\ee
where the last equality holds when the VEVs of the bi-doublet $\Phi$ are both real.  
The second realization requires the invariance w.r.t. $l_L \leftrightarrow 
l_L^c$, $\Delta_L \leftrightarrow \Delta^c_L\equiv \Delta_R$, 
$\Phi \leftrightarrow \Phi^c\equiv \Phi^T$, i.e. is essentially the charge 
conjugation symmetry. It leads to 
\be
f_L=f_R\equiv f\,,\qquad\quad y=y^T\,. 
\label{LR2a}
\ee
While both implementations of the discrete LR symmetry are possible, the 
second one  
is more natural in the context of $SO(10)$ unified theories, where it is an 
automatic gauge symmetry  \cite{Dsym}. In what follows we will adhere to this 
latter realization, i.e. will be assuming the equalities in eq. (\ref{LR2a}) 
to hold. The light neutrino mass matrix in this LR-symmetric seesaw mechanism 
is then given by 
\be
m_\nu \simeq m_\nu^{II}+m_\nu^{I} = v_L f- v^2\, y\, 
(v_R f)^{-1}\, y\,.
\label{ss2}
\ee
For the realization (\ref{LR1a}) of discrete LR symmetry, 
type I term in eq. (\ref{ss2}) would contain $(f^*)^{-1}$ rather than 
$f^{-1}$.

%%%%%%%%%%%%%%%%%%%%%%%%%%%%%%%%%%%%%%%%%%%%%

{\it 2. Seesaw and duality. } We will pursue a bottom-up approach: 
assuming $m_\nu$ and $y$ to be known, we solve eq.~(\ref{ss2}) for the 
$n\times n$ matrix $f$, where $n$ is the number of lepton generations.
Note that the matrices $m_\nu$ and $f$ are 
in general complex symmetric, whereas $y$ is symmetric due to the assumed 
discrete LR symmetry (\ref{LR2a}). 

An examination of eq. (\ref{ss2}) reveals the existence of its     
following duality property: if $f$ is a solution, so is 
\be
\hat{f}\equiv  \frac{m_\nu}{v_L} - f\,,
\label{dual1}
\ee
provided that the matrix $y$ is invertible. This can be readily verified by 
the direct substitution of (\ref{dual1}) into eq.~(\ref{ss2}). 
Notice that exactly the same duality holds in the 
case of the realization (\ref{LR1a}) of the discrete left-right symmetry.

The duality implies that, given $m_\nu$ and $y$, there is no unique solution 
for the triplet Yukawa coupling $f$. 
As we shall see, for $n$ fermion generations there are $2^n$ matrices $f$
that, for a given $y$, result in exactly the same mass matrix of light 
neutrinos $m_\nu$. It should be stressed that this duality is a unique 
property of the LR-symmetric type I+II seesaw: no such duality exists in the 
cases of pure type I or II seesaw, nor in the case when the two seesaw 
contributions are unrelated. 
 
In LR symmetric models with non-minimal particle content, the presence of 
extra particles may complicate the structure of the seesaw. In that case, in 
general there is no duality (at least in the form discussed in this paper). 

We shall now consider the illustrative examples of one and two lepton 
generations as well as the realistic three generations case.

%%%%%%%%%%%%%%%%%%%%%%%%%%%%%%%%%%%%%%%%%%%%%

{\it 3. One lepton generation}. In this case $m_\nu$, $y$ and $f$ are  
merely complex numbers. Eq. (\ref{ss2}) is a quadratic equation for $f$
with solutions 
\be
f_\pm=\frac{m_\nu}{2 v_L} \pm \sqrt{\frac{m^2_\nu}{4 v_L^2} + 
\frac{v^2y^2}{v_L v_R} }\,.
\label{exact}
\ee
The quantities $f_\pm$ satisfy $f_- + f_+ =m_\nu/v_L$, i.e. are duals of 
each other. In the limit 
$|m_\nu|^2 \gg 4|y^2 v^2 v_L/v_R|$, 
eq.~(\ref{exact}) yields 
\be
f_{-} \simeq -\frac{y^2 v^2}{m_\nu v_R}\,, \qquad\quad 
f_+\simeq \frac{m_\nu}{v_L}+\frac{y^2 v^2}{m_\nu v_R}\,. 
\label{approx1}
\ee
This limit leads to the dominance of  type I or II seesaw in the light neutrino 
mass, depending on which solution, $f_-$ or $f_+$, is realized in nature. 
For $v_L\ll |m_\nu|$, the solution $f_+$ violates the perturbative unitarity 
and must be discarded, so that only $f_-$ can be acceptable; a necessary 
condition for either solution to be physical is $|v_R| \gtrsim |v^2 y^2 
/m_\nu|$.

When the two terms under the square root in eq.~(\ref{exact}) are of the 
same order, the contributions of type I and type II terms to 
$m_\nu$ are also of the same order (mixed seesaw). Finally, 
in the limit $|m_\nu|^2 \ll 4|y^2 v^2 v_L/v_R|$, 
one has $m_\nu^I\simeq - m_\nu^{II}$ and $|m_\nu^{I,II}|\gg |m_\nu|$,
i.e. in this 
case there is a strong cancellation of the two terms in eq.~(\ref{ss2}). 

Consider now  $n$ fermion generations with no 
mixing (which means that $f$ and  $y$ are diagonal in the same basis). 
This case can be described as $n$ replicas of the one-generation case:
there exist $n$ pairs of solutions $f_{\pm}^{(i)}$ ($i=1,...,n$) to 
$n$ quadratic equations, and each solution for the matrix $f$ corresponds to 
picking one value $f_{+}^{(i)}$ or $f_{-}^{(i)}$ from each of $n$ pairs. The 
number of solutions is thus $2^n$. Switching on the mixing will change the 
nature of the solutions (which will no longer be diagonal), but not their 
number \cite{f1}.

%%%%%%%%%%%%%%%%%%%%%%%%%%%%%%%%%%%%%%%%%%%%%%

{\it 4. Two lepton generations}. In this case  $m_\nu$, $y$ and $f$ are  
symmetric $2\times 2$ complex matrices. For definiteness, we shall consider 
the $(\nu_\mu,\,\nu_\tau)$ sector of $m_\nu$. 
The matrices $f^{-1}$ and $y$ are given by 
\be
f^{-1}=\frac{1}{F}\left(\begin{array}{cc}
f_{33} & -f_{23}\\
-f_{23} & f_{22}
\end{array}\right)\,, \quad
y=\left(\begin{array}{cc}
y_{\mu2} & y_{\mu3}\\
y_{\tau2} & y_{\tau3}\end{array}\right)\,,
\label{f-1}
\ee
where $F\equiv det f$ and $y_{\tau2} = y_{\mu 3}$. 
Defining $x\equiv v_L v_R /v^2$, one can rewrite eq. (\ref{ss2}) in 
components as 
\bea
x F (f_{22}-m_{\mu\mu}) &=& f_{33} y_{\mu 2}^2-2 f_{23} y_{\mu 2} y_{\mu 3}
+f_{22} y_{\mu 3}^2\,, \nonumber \\
x F (f_{33}-m_{\tau\tau}) &=& f_{33} y_{\mu 3}^2-2 f_{23} y_{\mu3} 
y_{\tau 3}+f_{22} y_{\tau 3}^2 \,,
\nonumber \\
x F (f_{23}-m_{\mu\tau}) &=& f_{33} y_{\mu 2} y_{\mu 3}-f_{23} (y_{\mu 2} 
y_{\tau 3}+ y_{\mu 3}^2) \nonumber \\
& & + f_{22} y_{\mu 3} y_{\tau 3}\,, 
\label{sys1}
\eea
where $m\equiv m_\nu/v_L$. This is a system of coupled cubic equations for 
the matrix elements of $f$. Usually, solving systems of this kind presents 
serious difficulties; one has to resort to numerical methods, whose 
application is complicated by the fact that non-linear equations have 
multiple solutions, and their number is often not known in advance. We shall, 
however, show now that the system of equations (\ref{sys1}) admits a simple 
exact analytic solution. 
To find this solution, we develop here a method that may also prove to be 
useful for solving similar systems of coupled non-linear equations appearing 
in different contexts. 

Let us rescale all the matrices of interest according to 
\be
f_{ij}=\sqrt{\lambda} f_{ij}'\,,\quad
m_{ij}=\sqrt{\lambda} m_{ij}'\,,\quad
y_{ij}=\sqrt{\lambda} y_{ij}'\,,
\label{resc1}
\ee
where $\lambda$ is an as yet arbitrary complex number. The scaling law 
was chosen in such a way that in terms of the primed variables the system 
of equations for $f_{ij}'$ has the same form as eq.~(\ref{sys1}). Next, we 
fix the value of $\lambda$ by requiring $F'\equiv det f' = 1$.
The system of equations for $f_{ij}'$ then becomes linear and can be readily 
solved. Expressing the primed variables 
back through the unprimed ones 
and substituting them into the condition $F'=1$, one obtains a 4th order 
characteristic equation for $\lambda$. Solving it completes the solution of 
the problem.

Since the matrix $y$ is symmetric, one can go, without a loss of generality,  
to the basis where $y$ is diagonal: $y=diag(y_{2},\,y_{3})$. 
The solution for the matrix $f$ then takes the form 
\be
f=\dfrac{x\lambda}{(x\lambda)^2-y_2^2 y_3^2} \!\left(\begin{array}{cc} 
\!\! x\lambda m_{\mu\mu} + y_2^2 m_{\tau\tau} & 
m_{\mu\tau} (x\lambda-y_2 y_3) \\
\!\!\dots & x\lambda m_{\tau\tau} + y_3^2 m_{\mu\mu}
\end{array}\!\! \right),
\label{sol2}\eeq
where $\lambda$ has to be found from the characteristic equation 
\bea
\left[(x\lambda)^2 - y_2^2 y_3^2\right]^2 - x\left[\det m (x\lambda - 
y_2 y_3)^2 x\lambda \right.\nonumber \\
\left. \qquad\qquad + (m_{\mu\mu} y_3 + m_{\tau\tau} y_2)^2 (x\lambda)^2 
\right] = 0\,.
\label{lam2}
\eea
Eq.~(\ref{lam2}) is quartic in $\lambda$ and so has in general four complex 
solutions, leading to four solutions for the matrix $f$, as expected. 

Rewriting eq. (\ref{ss2}) as $x \hat{f} = - y f^{-1} y$ and taking the determinants of both sides, one finds 
$x^2 F \hat{F} = y_2^2 y_3^2$,
where $\hat{F}\equiv \det \hat{f}$. 
Since the condition $F'=1$ corresponds to $F=\lambda$, this equation
leads to $x^2 \lambda \hat{\lambda} = y_2^2 y_3^2$.
With the help of this relation, 
it is straightforward to check that the four structures of $f$ defined by 
eqs.~(\ref{sol2}) and (\ref{lam2}) form two dual pairs, and  
it is also easy to express the four solutions of 
eq.~(\ref{lam2}) in a closed form:
\beq
x\lambda_i = \frac{1}{4}[x \det m + r_{\pm} \pm \sqrt{2(\det m)^2 x^2 + 
4 k x + 2 x r_\pm \det m }],
\label{solu}\eeq
where
\bea
r_\pm &=& \pm \sqrt{(\det m)^2 x^2+4 k x + 16 y_2^2 y_3^2}\,,
\nonumber \\
k &=& m_{\mu\mu}^2 y_3^2+2 m_{\mu\tau}^2 y_2 y_3+m_{\tau\tau}^2 y_2^2\,.
\label{k}
\eea
Eq.~(\ref{solu}) gives two pairs of dual 
solutions, one of them corresponding to $r_+$ and another to $r_-$. In 
each dual pair the solutions differ by the sign in front of the radical.

{}From eq.~(\ref{sol2}) one can see that when a solution $\lambda_1$ 
satisfies $|x\lambda_1| \gg |y_i y_j|$ ($i,j=2,3$), one obtains 
$f_1 \simeq m$, which corresponds to type II seesaw case. Then the  
dual solution $x\lambda_2=x\hat{\lambda}_1=y_2^2 y_3^2/(x\lambda_1)$ 
has modulus $\ll |y_i y_j|$ 
and the corresponding matrix $f_2=\hat{f}_1$ takes the form obtained in 
type I seesaw case. Notice that eqs. (\ref{solu}) and (\ref{k}) imply that  
the condition $|x\lambda_1| \gg |y_i y_j|$ can only be satisfied in the 
limit $|m_{\alpha\beta} m_{\gamma\delta}| \gg 4|y_i y_j /x|$ 
($\alpha,\beta,\gamma,\delta = \mu,\tau$). This condition ensures the 
existence of solutions with the dominance of one seesaw type, in analogy 
with the one generation case. However, in general not all four solutions 
correspond to the one seesaw type dominance in this limit:
if $|\det m| \gg 4|y_i y_j /x|$, the remaining two dual solutions $f_3$ and 
$f_4$ are of mixed type.
When $|m_{\alpha\beta}m_{\gamma\delta}| 
\lesssim 4|y_i y_j /x|$, in general all four solutions are of mixed type.
A similar classification applies to the 3-generation case.

%%%%%%%%%%%%%%%%%%%%%%%%%%%%%%%%%%%%%%%%%%%%%

{\it 5. Three lepton generations}. In this case $f$, $y$ and $m=m_\nu/v_L$ 
are complex symmetric $3\times 3$ matrices. As in the 2-generation case, 
we go to the basis where $y$ is diagonal: $y=diag(y_1,\,y_2,\,y_3)$. 
Quite analogously to the derivation of eq.~(\ref{sys1}), one can obtain  
from (\ref{ss2}) equations for the matrix elements of $f$ and its dual 
$\hat{f}$: 
\begin{eqnarray}
xF(f_{ij}-m_{ij})&=&y_i y_j F_{ij} ~,\label{eqsf}\\
x\hat{F}(\hat{f}_{ij}-m_{ij})&=&-x\hat{F}f_{ij}=y_i y_j \hat{F}_{ij}~,
\label{eqsft}
\end{eqnarray}
where $F\equiv \det{f}$, $\hat{F}\equiv\det{\hat{f}}$ as before, and 
\beq
\begin{array}{ll}
F_{ij}\equiv \frac 12 \epsilon_{ikl}\epsilon_{jmn} f_{km}f_{ln} ~,
\nonumber \\
\hat{F}_{ij}\equiv \frac 12 \epsilon_{ikl}\epsilon_{jmn} 
\hat{f}_{km}\hat{f}_{ln}
= M_{ij}-T_{ij}+F_{ij} ~,\\
M_{ij}\equiv \frac 12 \epsilon_{ikl}\epsilon_{jmn} m_{km}m_{ln} ~,~~ 
T_{ij}\equiv \epsilon_{ikl}\epsilon_{jmn} f_{km} m_{ln} ~.
\end{array}
\label{tensor}
\eeq
Taking the determinants of both sides of the equation $x\hat{f}= -yf^{-1}y$, 
one obtains 
\beq
x^3 F \hat{F} = - y_1^2 y_2^2 y_3^2\,.
\label{du3}
\eeq

Eq.~(\ref{eqsf}) is a system of six coupled quartic equations for the 
elements $f_{ij}$ (note that $F$ is cubic in $f_{ij}$ in the 3-generation 
case). Since the right-hand sides of these equations are quadratic rather 
than linear in $f_{ij}$, a simple rescaling would not immediately linearize 
the system. However, using the dual system of equations (\ref{eqsft}) one 
can write for these right-hand sides $y_i y_j F_{ij} = -x\hat{F}f_{ij} + 
y_iy_j(T_{ij} - M_{ij})$. 
The resulting system of equations can now be linearized by a rescaling, 
similar to that in eq.~(\ref{resc1}), except that the scaling factor is now 
$\lambda^{1/3}$ rather than $\sqrt{\lambda}$. Once again, we fix $\lambda$ 
by requiring $F'(\lambda)=1$; eq.~(\ref{du3}) then yields 
$\hat{F}'=-(y_1' y_2' y_3')^2/x^3$. The linearized system is 
\beq
[x^3-(y'_1 y'_2 y'_3)^2] f'_{ij} - x^3m'_{ij}= x^2 y'_i y'_j 
(T'_{ij}-M'_{ij})\,. 
\label{full}
\eeq
The characteristic equation for $\lambda$ is now of 8th order and its 
solutions lead to four pairs of dual solutions $f$ and $\hat{f}$ for the 
system (\ref{eqsf}). This 8th order equation and the resulting general 
analytic solution for $f$ are rather lengthy and will be given elsewhere; 
here we present the simplified case $y_1\to 0$. 

The case $|y_1|\ll |y_{2,3}|$ is actually physically well motivated, in view 
of the smallness of the masses of first-generation charged fermions. Notice 
that for $y_1=0$ the matrix $y$ is not invertible, so that the proof of the 
duality of solutions for $f$ 
breaks down. However, one can check that the solutions of eq.~(\ref{eqsf}) 
for $y_1=0$ are the same as the ones obtained by setting $y_1=0$ in 
eq.~(\ref{full}). In other words, one can safely use the duality and take 
the limit $y_1\rightarrow 0$ in the final result. The solution for the 
symmetric matrix $f$ is then given in components as 
\bea
f_{11}=m_{ee}\,,\qquad 
f_{12}=m_{e\mu}\,,\qquad f_{13}=m_{e\tau}\,, \nonumber \\
f_{23}=\left(m_{\mu\tau}+\dfrac{y_2y_3 m_{e\mu} 
m_{e\tau}}{x\lambda}\right)/d_2 
\,,\qquad \nonumber
\eea
\bea
f_{22}=\left[m_{\mu\mu} + \dfrac{y_2^2}{x \lambda} \left(M_{22}-
\dfrac{y_3^2 m_{ee} m_{e\mu}^2}{x\lambda}\right)\right]/d_1\,, 
\nonumber \\
f_{33}=\left[m_{\tau\tau} + \dfrac{y_3^2}{x\lambda} 
\left(M_{33}-\dfrac{y_2^2 m_{ee} m_{e\tau}^2}{x\lambda}\right)\right]/d_1\,,
\label{big}
\eea
where
\be
d_1=1 - \dfrac{y_2^2 y_3^2 m_{ee}^2 }{(x \lambda)^2}\,,\qquad\quad
d_2=1 + \dfrac{y_2y_3m_{ee}}{x\lambda}\,.
\label{dd}
\ee

In the considered limit $y_1\to 0$ the characteristic equation for $\lambda$ 
reduces to 
\bea
\lambda^4 \left\{[(x\lambda)^2 -  m_{ee}^2y_2^2 y_3^2]^2
- x \left[\det m (x\lambda - m_{ee}y_2 y_3 )^2 x\lambda \right.\right.
\nonumber \\
\left.\left. 
+ (M_{22} y_2 + M_{33} y_3)^2 (x\lambda)^2\right]\right\} = 0\,.~~~
\label{lam32}
\eea

A comment on the duality and multiplicity of solutions in the 3-generation 
case is in order. If a solution $\lambda$ of the general characteristic 
equation is non-zero in the limit $y_1\rightarrow 0$, then $\hat{\lambda}=
-y_1^2 y_2^2 y_3^2/ (x^3\lambda) \rightarrow 0$, which means that the 
determinant of $\hat{f}$ vanishes. Thus, the dual of any solution that 
is finite for $y_1\to 0$ becomes singular and must be discarded. Therefore, 
in this limit there are only four (rather than eight) solutions with no 
duals. The corresponding values of $\lambda$ are the zeros of the factor in 
curly brackets in eq.~(\ref{lam32}), which is quartic in $\lambda$. From the 
comparison with eq.~(\ref{lam2}) a strong connection with the pure 
2-generation case is evident, so that a {\sl different duality} among the 
four remaining solutions exists: if $\lambda$ satisfies eq.~(\ref{lam32}), 
so does $\tilde{\lambda}\equiv y_2^2 y_3^2 m_{ee}^2 /(x^2 \lambda)$, and it 
corresponds to $\tilde{f}\equiv \tilde{m}-f$, where $\tilde{m}_{\alpha\beta}
=m_{\alpha\beta}+m_{e\alpha} m_{e\beta}/m_{ee}$. There are two pairs of 
such solutions.

%%%%%%%%%%%%%%%%%%%%%%%%%%%%%%%%%%%%%%%%%%%%%

{\it 6. Concluding remarks.} We have revealed and analyzed an interesting 
duality property of the seesaw mechanism with contributions of both 
RH neutrinos and isotriplet Higgs bosons 
in the presence 
of discrete left-right symmetry. In particular, it has been shown that 
for a given mass matrix of light neutrinos $m_\nu$ and Dirac-type Yukawa 
coupling matrix $y$, there are multiple solutions for the Majorana-type 
Yukawa coupling matrix $f$ and thus for the matrix of RH neutrinos $M_R$. 
The number of solutions, however, does not exceed eight (for three lepton 
generations), and with the help of the developed here formalism they can be 
readily analyzed one by one. 
In contrast to this, in models where type I and type II contributions to 
neutrino mass are unrelated, 
there are in general infinitely many possible decompositions of $m_\nu$ 
into these two parts.

The discrete LR symmetry of the underlying theory must be broken 
at some scale $v_{LR}$, and the renormalization group evolution below this 
scale may result in a violation of conditions (\ref{LR2a}) at lower energies.
The corrections to the matrix elements of $y$ and $f$ depend logarithmically 
on the ratios of the masses of RH neutrinos and  Higgs triplets and are 
suppressed by loop factors and possibly by small couplings. We checked 
numerically that the reconstruction of the matrix $f$ as performed in this 
paper remains accurate at a percent level if the LR-violating corrections 
to the matrix elements are of the order of percent. 
This stability, however, may be lost if small matrix elements receive 
corrections proportional to the large ones. This issue requires a 
dedicated study, which is beyond the scope of the present paper.

Possible applications of our results include the bottom-up reconstruction of 
the mass matrix of heavy RH neutrinos and model building. The bottom-up 
approach requires the knowledge of the light neutrino mass matrix and of 
the Dirac-type Yukawa couplings $y$. The former is partially known from the 
experiment (absolute mass scale, 1 - 3 mixing angle and CP 
violating phases have not been measured yet), whereas for $y$ one needs
both data and additional theoretical assumptions, such as quark-lepton 
symmetry or Grand Unification. Model builders can use our results to 
immediately find out the allowed structures of the triplet Yukawa couplings 
$f$ that can successfully reproduce the low energy phenomenology, which  
may help them to systematically look for the underlying symmetries of the 
mechanism of lepton mass generation.

We thank R. N. Mohapatra, V. A. Rubakov, G. Senjanovi\'c and especially 
A. Yu. Smirnov for useful discussions. 
This work was supported by SFB-375 f\"ur Astro-Teilchenphysik der Deutschen 
Forschungsgemeinschaft (EA) and by U.S. Department of Energy grant no. DE-FG03-94ER40837 (MF). 

%%%%%%%%%%%%%%%%%%%%%%%%%%%%%%%%%%%%%%%%%%%%%

\end{document}